# ELECTRON SCATTERING BY SHORT RANGE DEFECTS AND RESISTIVITY OF GRAPHENE

Natalie E. Firsova*†, Sergey A. Ktitorov*†

*The Ioffe Physical-Technical Institute of the Russian Academy of Sciences, 26 Politekhnicheskaya, St. Petersburg 194021, Russia

†Peter the Great St. Petersburg Polytechnic University, 29 Polytechnicheskaya, St. Petersburg 195251, Russia

†Saint Petersburg Electrotechnical University, 5 Professora Popova, 197376 St. Petersburg, Russia

## ABSTRACT

The electron scattering by the short-range defects in the monolayer graphene is considered in the framework of the flatland model. We analyze the effect of this scattering on the electronic resistivity of the monolayer graphene (direct problem) and develop a procedure for determination of the defect potential from resistivity data (inverse problem). We use an approximation of the short-range perturbation by the delta-shell potential that is reasonable since it suppresses irrelevant short wavelength excitations. Our theoretical results proved to be in a good agreement with experiment on suspended monolayer graphene. It means that our model correctly describes essential features of the physical problem under consideration. It gives possibility to consider the inverse problem, i.e. on the basis of our results for direct problem to develop a procedure for determination of parameters of the monolayer graphene sample using experimental measurements for it. Thus the obtained results give new important possibilities, which can be used in numerous applications.

## I. INTRODUCTION

Unique electronic properties of graphene attracted great attention of researchers from the beginning due to specific features of the electron and phonon systems and to prospects of applications. Quasi relativistic zero-gap electronic spectrum in the vicinity of the so-called Dirac points permits to construct the conductivity theory on the basis of the relativistic scattering theory [1].

In the present paper we analyze the short-range defects effect on the monolayer graphene resistivity in the framework of flatland model. We use the scattering theory for the two-dimensional Dirac equation developed by D.S. Novikov [2], and the Born approximation. We approximate the short range defects by the suggested in [3-5] delta-shell potential taking into account a local chemical potential perturbation arising due to the defect presence. This model allows us to suppress irrelevant short wavelength excitations, which seems to be reasonable.

* Corresponding author
E-mail: ktitorov@mail.ioffe.ru

Our goal in this work is to continue the study of the resistivity in monolayer graphene within the framework of flatland model and delta shell potential approximation. Moderately near the Dirac point we use the formula for conductivity obtained in [6] on the basis of the partial wave technique and we calculate below the resistivity (conductivity) in the wide range of energies using the Born approximation. We compare our theoretical results with the available experimental data on the suspended monolayer graphene [7].

Dirac equation describing electronic states in zero-gap graphene reads [1]

$$\hat{\mathrm{H}}\psi(\mathrm{x},\mathrm{y}) \equiv \left[-i\hbar \mathrm{v}_F \sum_{\mu=1}^{2} \hat{\sigma}_\mu \partial_{\mathrm{x}_\mu} + \hat{\mathrm{V}}\right]\psi(\mathrm{x},\mathrm{y}) = \mathrm{E}\psi(\mathrm{x},\mathrm{y}), \tag{1}$$

where $\mathrm{v}_F$ is the Fermi velocity, $\sigma_\mu$ are the Pauli matrices, $\psi(\mathbf{r})$ is the two-component spinor.

We use the delta - shell potential model [3-5] approximating the actual short-range perturbation by:

$$\mathrm{V}(\mathrm{r}) = -\mathrm{V}_0\,\delta(\mathrm{r}-\mathrm{r}_0).$$

.

It means that the perturbation is concentrated on the circumference of the circle that allows us to suppress the irrelevant deep atomic states that provides a regulation of the theory. Besides, we assume that perturbation locally shifts the Dirac point on the energy scale.

Let us introduce dimensionless variables:

$$\hat{H} = \frac{\hat{\mathrm{H}}}{\hbar \mathrm{v}_\mathrm{F}/\mathrm{r}_0},\quad E = \frac{\mathrm{E}}{\hbar \mathrm{v}_F/\mathrm{r}_0},\quad V_0 = \frac{\mathrm{V}_0}{\hbar \mathrm{v}_F},$$
$$x = \mathrm{x}/\mathrm{r}_0,\quad y = \mathrm{y}/\mathrm{r}_0,\quad r = \mathrm{r}/\mathrm{r}_0,\quad k = \mathrm{k}\,\mathrm{r}_0. \tag{2}$$

Thus, the dimensionless potential reads:

$$V(r) = -V_0\delta(r-1). \tag{3}$$

The equation (1) takes the form:

$$\hat{H}\psi(x,y) \equiv \left[-i\sum_{\mu=1}^{2}\hat{\sigma}_\mu \partial_{x_\mu} + \hat{V}\right]\psi(x,y) = E\psi(x,y). \tag{4}$$

## II. SCATTERING MATRIX: PARTIAL WAVES REPRESENTATION

Calculating the ratio of the incoming and outgoing waves, we have components of the S-matrix in the partial waves representation [4]:

$$S_j(k) = -\frac{F_j^{(2)}(k)}{F_j^{(1)}(k)}, \quad j = \pm 1/2, \pm 3/2, \quad k = E, \tag{5}$$

where

$$F_j^{(\alpha)} = \left[ I_{j-1/2}(k) H_{j+1/2}^{(\alpha)}(k) - I_{j+1/2}(k) H_{j-1/2}^{(\alpha)}(k) \right] \bullet$$
$$\tan V_0 \left[ I_{j+1/2}(k) H_{j+1/2}^{(\alpha)}(k) + I_{j-1/2}(k) H_{j-1/2}^{(\alpha)}(k) \right], \; \alpha = 1, 2. \tag{6}$$

Asymptotically exact expression for the transport relaxation time for arbitrary $V_0$ and $k \to 0$ reads [6]:

$$\tau_{tr}^{-1}(\mathrm{k}) = \pi^2 \, \mathrm{kv}_F \, N_i \tan^2 V_0, \tag{7}$$

where $N_i = \mathrm{r}_0^2 \, \mathrm{N}_i$, $\mathrm{N}_i$ is the concentration of defects. Assuming the Boltzmann equation to be valid, we can write the following formula for the conductivity at low temperature $\mathrm{k_B T} << \mathrm{E_F}$:

$$\sigma(\mathrm{E_F}) = \sigma_0 \frac{\mathrm{E_F} \, \tau_{tr}(\mathrm{E_F})}{\hbar}, \quad \tilde{\sigma} = \sigma / \sigma_0, \tag{8}$$

where $\sigma_0 = \frac{4e^2}{h}$, the factor 4 takes into account spin and valley degeneracy, $h = 2\pi\hbar$ is the

Planck constant. The relaxation time $\tau_{tr}(\mathrm{E_F})$ and the conductivity $\sigma(\mathrm{E_F})$ were calculated in [6]. Asymptotically exactly in the low Fermi energy limit**.** Therefore, the Boltzmannian resistivity $\rho$ reads [6]

$$\rho = \sigma^{-1} = \sigma_0^{-1} \pi^2 N_i \cdot \tan^2 V_0,$$
$$\tilde{\rho} = \rho / \rho_0, \; \rho_0 = \sigma_0^{-1}, \tag{9}$$

where $\tilde{\rho}$ is the dimensionless resistivity**.** Here we presented the formula for resistivity, which is correct in the long wavelength limit and for arbitrary value of the perturbation within the range, where the Boltzmann equation is valid:

$$\mathrm{E}_F\, \tau_{tr} / \hbar > 1. \tag{10}$$

Below we calculate the resistivity for a wide range of wavelengths, but in the limit of small potential. Thereto we apply the Born approximation [2, 5].

## III. BORN APPROXIMATION

The Dirac equation (1) can be written in the integral form [2, 5]:

$$\psi_{scat} = -\int dx'dy' G\left(x-x', y-y'\right)\left[-i\hat{\sigma}_\mu \partial_{x_\mu} + E\right]\hat{V}\left(x', y'\right) u_p e^{i\mathbf{pr}'}, \tag{11}$$

where the Green function reads:

$$G\left(\mathbf{r}\right) = \int \frac{dk_x dk_y}{\left(2\pi\right)^2} \frac{e^{i\left(k_x x + k_y y\right)}}{k^2 - \left(E + i0\,\mathrm{sgn}\,E\right)^2} = \frac{i\pi}{4\pi}\mathrm{sgn}\,E \cdot H_0^{(1)}\left(kr\right),$$

$$H_m^{(1,2)}\left(x\right) = J_m\left(x\right) \pm iN_m\left(x\right), \;\; r = \sqrt{x^2 + y^2}$$

The equation (11) is written in the first Born approximation. The solution can be written as follows

$$\psi = u_{kx} e^{ikx} + \frac{f\left(\theta\right)}{\sqrt{-ir}} u_{p\theta} e^{ikr}, \tag{12}$$

where the scattering amplitude reads

$$f^{\,\mathrm{Born}}\left(k, \theta\right) = -\sqrt{k/8\pi} V_q \cdot \left(1 + e^{i\theta}\right), \tag{13}$$

and

$$\theta = \angle\left(\mathbf{k}', \mathbf{k}\right), \;\; q = 2\kappa \sin\left(\theta/2\right).$$

The Fourier transform reads

$$V_q = \int dxdy e^{-i\left(q_x x + q_y y\right)} V\left(r\right). \tag{14}$$

Let us introduce the notation: $\tilde{\Sigma}_{tr} = \Sigma_{tr} / \mathrm{r}_0$, where $\Sigma_{tr}$ is the transport crosssection. Substituting equation (14) into (13) and making use of the well known formula for the dimensionless transport cross section [2]

$$\tilde{\Sigma}_{tr} = \int d\theta (1 - cos\theta) \left| f(\theta) \right|^2 ,$$

we obtain the dimensionless transport cross section for our problem in the form:

$$\tilde{\Sigma}_{tr}^{Born}(k) = \pi k V_0^2 I(k). \quad (15)$$

Here the function $I(k)$ is given by the formula:

$$I(k) = \int_0^{2\pi} d\theta \sin^2(\theta) J_0^2 \left[ 2k \sin(\theta/2) \right] = 4 \int_0^{\pi/2} d\vartheta \sin^2(4\vartheta) J_0^2 \left[ 2k \sin(2\vartheta) \right],$$

where $\theta = 4\vartheta$.

(16)

This function can be expressed in terms of the hypergeometric functions [8]:

$$I(k) = \frac{1}{2} \Gamma \begin{bmatrix} 3/2,\ 1/2 \\ 1,\ 1,\ 2 \end{bmatrix} \cdot {}_4F_5 \left( 3/2,\ 1/2,\ 1/2,\ 1;\ 1,\ 1,\ 1,\ 3/2;\ -k^2 \right). \quad (17)$$

Here

$$\Gamma \begin{bmatrix} \alpha_1, \alpha_2 \\ \beta_1,\ \beta_2,\ \beta_3 \end{bmatrix} = \frac{\Gamma(\alpha_1)\Gamma(\alpha_2)}{\Gamma(\beta_1)\Gamma(\beta_2)\Gamma(\beta_3)}, \quad (18)$$

$\Gamma(\alpha)$ is the gamma function, ${}_4F_5$ is the generalized hypergeometric function:

$${}_4F_5(\alpha_1, \alpha_2, \alpha_3, \alpha_4; \beta_1, \beta_2, \beta_3, \beta_4, \beta_5;\ z) = \sum_{n=0}^{\infty} \frac{(\alpha_1)_n (\alpha_2)_n (\alpha_3)_n (\alpha_4)_n}{(\beta_1)_n (\beta_2)_n (\beta_3)_n (\beta_4)_n (\beta_5)_n} \frac{z^n}{n!},$$

where $(\alpha)_n = \dfrac{\Gamma(\alpha + n)}{\Gamma(n)}$ is the rising Pohhammer symbol.

Using (15), we find the relaxation time from the well known formula

$$\frac{1}{\tau_{tr}^{Born}} = \Sigma_{tr}^{Born} \mathrm{N}_i \mathrm{v}_\mathrm{F} = \tilde{\Sigma}_{tr}^{Born} \mathrm{r}_0 \mathrm{N}_i \mathrm{v}_\mathrm{F} =$$
$$\pi k_i V_0^2 \mathrm{r}_0 \mathrm{N}_i \mathrm{v}_\mathrm{F} I(k) = \pi N_i V_0^2 \mathrm{k}_F \mathrm{v}_F I(k), \quad (19)$$

Substituting (19) into (8, 9), we obtain a formula for the dimensionless resistivity:

$$\tilde{\rho}^{Born} = \pi N_i V_0^2 I(k). \tag{20}$$

## IV. RELATION BETWEEN THE ELECTRONS SCATTERING DATA AND RESISTIVITY

We see that in the limit of $k \to 0$ we have $I(k) \to \pi$. Thus the expressions (9) and (20) for the dimensionless resistivity are asymptotically identical in the limit of $V_0 \to 0,\ k \to 0,$ i. e. in the vicinity of the resistivity maximum.

Numerical calculation of the functions (9) and (20) for resistivity allows us to compare the theoretical curve $\rho(E_F)$ (Figs. 1, 2) and the measured in the experiment on the suspended graphene [7] (Figs. 3, 4) resistivity near the Dirac point. The comparison shows rather near similarity of these curves shapes that indicates a possibility to establish useful relations between the scattering data and resistivity. There are two distinct domains of the Fermi energy, where a comparison with the experimental data can be carried out effectively: (i) in the vicinity of the resistivity maximum and (ii) on the slope of the resistivity contour. Since our theory takes into account only scattering on neutral impurities, we consider only experiment on annealed suspended graphene containig minimal amount of charged defects [7]. The shape of the $\rho(E_F)$ obtained in [7] can be well described by the function $I(E_F)$(see Fig. 1). The power function behavior of $\rho(E_F)$ on the slope is universal and follows from the short-range interaction of electrons with the defect approximated by the delta shell potential. Log – log graph of this function presented on Fig. 2 indicates with good accuracy the power exponent equal to – 1.

Notice that the experimental curve (Figs. 3, 4) deviates from the $1/E$ law at lower values of $E_F$ that can be traced back to a presence of distinct scatterers.

On the contrary, a behavior in the vicinity of maximum is essentially non-universal depending on the sample. The reason of it is the Dirac singularity. Different approaches give different values for the maximal possible theoretical resistivity [9, 10]. For instance, in [9] results do not depend on scattering data, because resistivity there has a diffractional nature i.e. de Broglier waves diffraction as in the Landauer formula. The maximal possible resistivity there equals

$$\frac{h}{4e^2}\eta^{-1}, \tag{21}$$

where $\eta$ is a numerical factor, its value depending on the order of limit $\omega \to 0,\ E_F \to 0,\ T \to 0.$

According to [9], good agreement with experiment is given by $\eta = \pi/4.$ Then the maximal resistivity $\rho_{trans}^{(1)}$ reads (notation by Ziegler [9]):

$$\rho^{(1)}{}_{max} = \frac{h}{\pi e^2} = \frac{24}{\pi}\ \text{k}\Omega \approx 7.64\ \text{k}\Omega\ . \tag{22}$$

In [10] the optical minimal conductivity is obtained on the basis of a solution of the von Neumann equation for the density matrix. It is found there that $\rho_{\max}^{opt} = \frac{h}{\pi^2 e^2}$.

On the opposite to [9, 10], our formula is based on the scattering data. The minimal conductivity we find from the condition of applicability of the Boltzmann kinetic equation, i.e. the Mott condition that the electron wavelength cannot be more, than the mean free path.We introduce a quasi-classical definition $\rho_{\max}^{qc}$ for maximum possible resistivity basing on the arguments by Mott: that the mean free path cannot be less, than the electron wavelength. Therefore, $\rho_{\max}^{qc}$ is determined by the condition:

$$E_F \tau_{tr} / \hbar = k_F l > 1, \tag{23}$$

where l is the mean free path. Then we have the following estimate for $\rho_{\max}^{qc}$:

$$\rho_{\max}^{qc} = h / 4e^2 \approx 6 \text{ k}\Omega. \tag{24}$$

For the comparison, measurements on the suspended graphene carried out in [7] gave in result $\rho_{\max}^{\exp} \approx 2.4 \text{ k}\Omega$, i. e. by 2.5 times less value. This shows that quantum corrections are important near the maximum.

Let us introduce the notion of scattering intensity $P$:

$$P = N_i r_0^2 \tan^2 \frac{V_0}{\hbar v_F} = N_i \tan^2 V_0. \tag{25}$$

For $V_0 = \frac{V_0}{\hbar v_F} << 1$ we have

$$P = N_i r_0^2 \left( \frac{V_0}{\hbar v_F} \right)^2 = N_i V_0^2. \tag{26}$$

Then formulae (9), (20) can be rewritten for $V_0 << \hbar v_F$ as follows:

$$P = \tilde{\rho}_{\max} / \pi^2. \tag{27}$$

Let us rewrite the obtained formula for $\tilde{\rho}_{\max}$ in the dimensional form

$$\rho_{\max} = \frac{h}{4e^2} \pi^2 P = \frac{h}{4e^2} N_i r_0^2 \tan^2 \frac{V_0}{\hbar v_F} \tag{28}$$

or for $V_0 << \hbar v_F$:

$$\rho_{\max} \approx \frac{h}{4e^2} N_i r_0^2 \left( \frac{V_0}{\hbar v_F} \right)^2 . \qquad (29)$$

Formulae (28), (29) give possibility to obtain information about parameters of defects (for instance, scattering intensity) basing on such experimental data as a resistivity for small Fermi energy.

## V. CONCLUSIONS

In the present paper we studied the electron scattering by short-range defects in the monolayer graphene and its influence on the resistivity (conductivity). We considered this problem in the framework of flatland model approximating short-range defects by the delta-shell potential. This form of the short-range potential secures regularization of the scattering problem due to suppression of irrelevant short wavelength excitations. In the low-energy limit we used the conductivity dependence on the Fermi level position obtained in [6] by partial wave technique. For this goal in [6] the corresponding 2+1-dimensional Dirac equation was solved and the asymptotic analysis of the found S - matrix allowed us to obtain the formula for conductivity we used.

The case of higher energies we studied using the Born approximation. Numerical calculations were carried out in the wide range of electronic energies E. The resistivity (conductivity) was shown to be a constant near the Dirac point and then to behave as 1/E if the energy was not too large because for high values of energy the Dirac equation is not valid (see Figs.1, 2).

We compared our theoretical curves (Figs. 1, 2) and the experimental measurements made in [7] for suspended annealed monolayer graphene (Figs. 3, 4). From this comparison we concluded that the considered model satisfactory described the resistivity of the suspended graphene due to the minimization of the amount of charged defects in it in result of annealing. Comparison of our theoretical results with the experimental data indicates that our theory is in good agreement with experiment on suspended monolayer graphene for moderately low Fermi energy.

Comparing our theoretical result (Figs 1, 2) with the experimental data (Figs. 3, 4) makes us conclude that the experimental curve can be described by the $1/E$ law at lower energies, than it is predicted by our theory, in result of action of an alternative scattering mechanism.

Our theory being applied to analysis of the experiments on suspended monolayer graphene gives a tool for good estimate of the scattering intensity of the studied sample if we measured its maximal value of resistivity near the Dirac point (for low values of Fermi energy) . Thus, the obtained results are very important for applications. One of the most promising applications in electronics is the planar field-effect transistor. Colossal electronic mobility of the monolayer graphene makes it prospective for constructing high-frequency devices.

## FIGURE CAPTIONS

Fig. 1. Born scattering amplitude as a function of the Fermi energy (dimensionless variables).

Fig. 2. Log – log graph for the Born scattering amplitude as a function of the Fermi energy.

Fig. 3. Measured resistivity as a function of the Fermi energy.

Fig. 4. Log – log graph for the measured resistivity as a function of the Fermi energy.

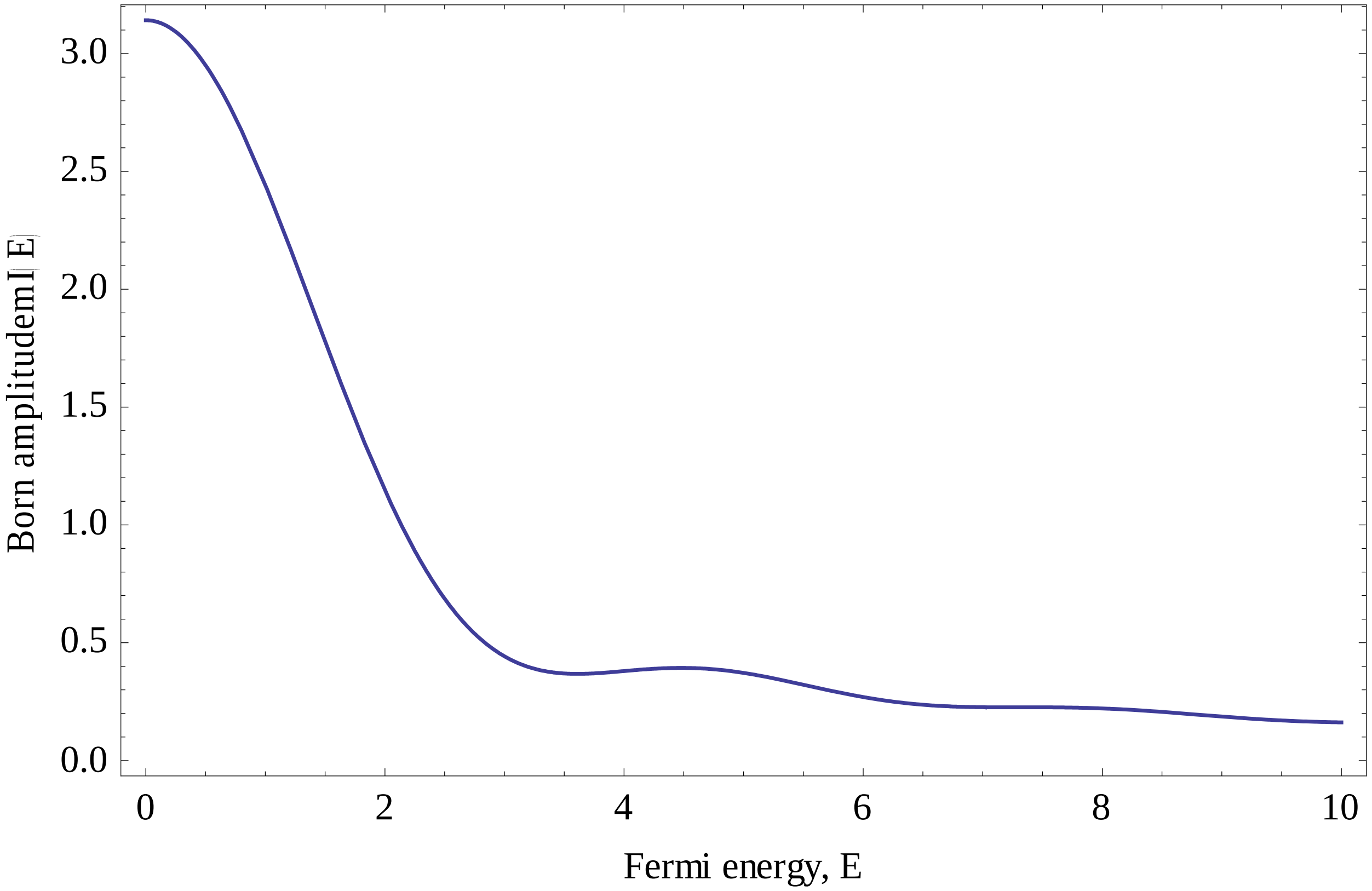

**Fig. 1**

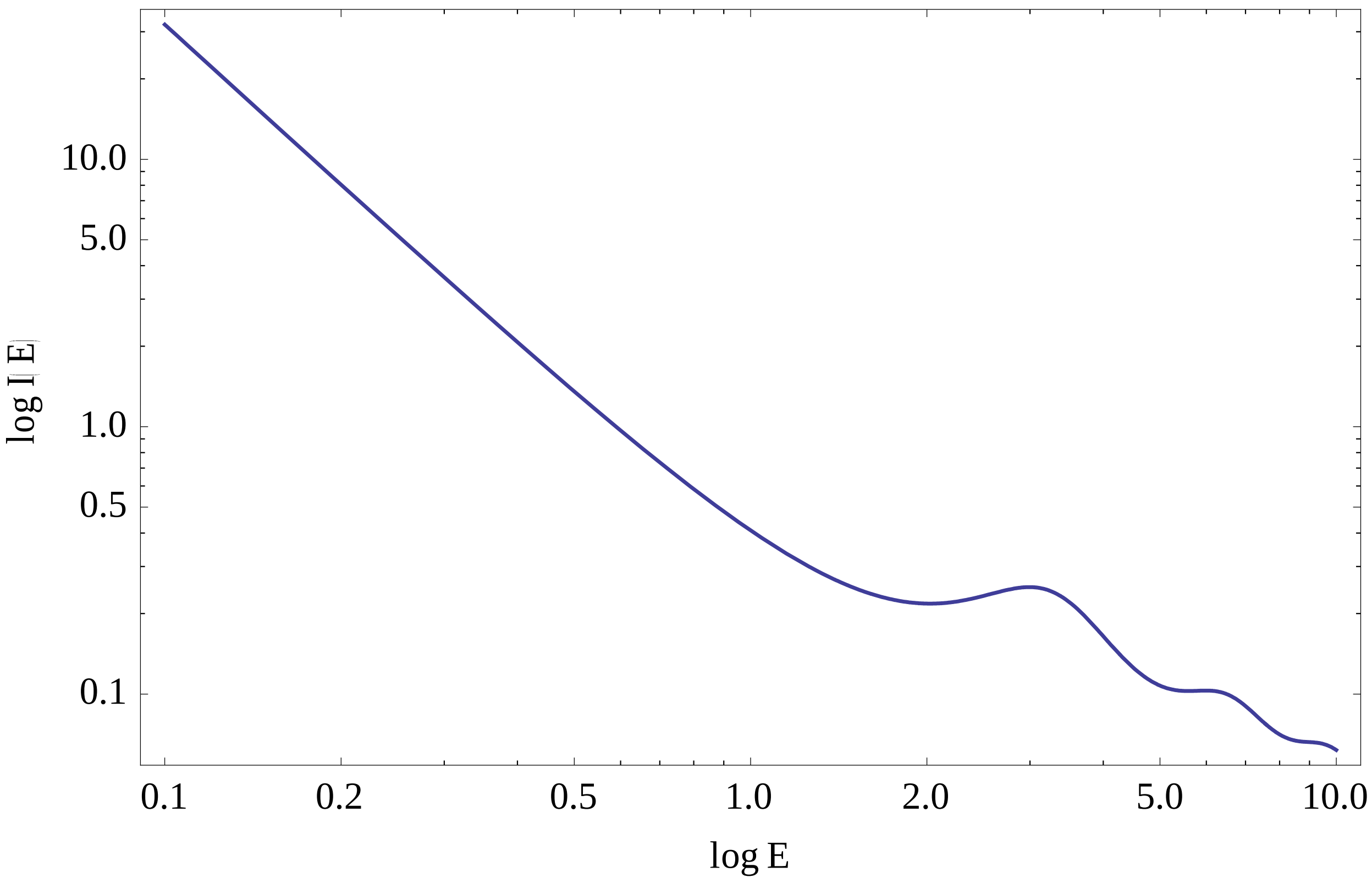

**Fig. 2**

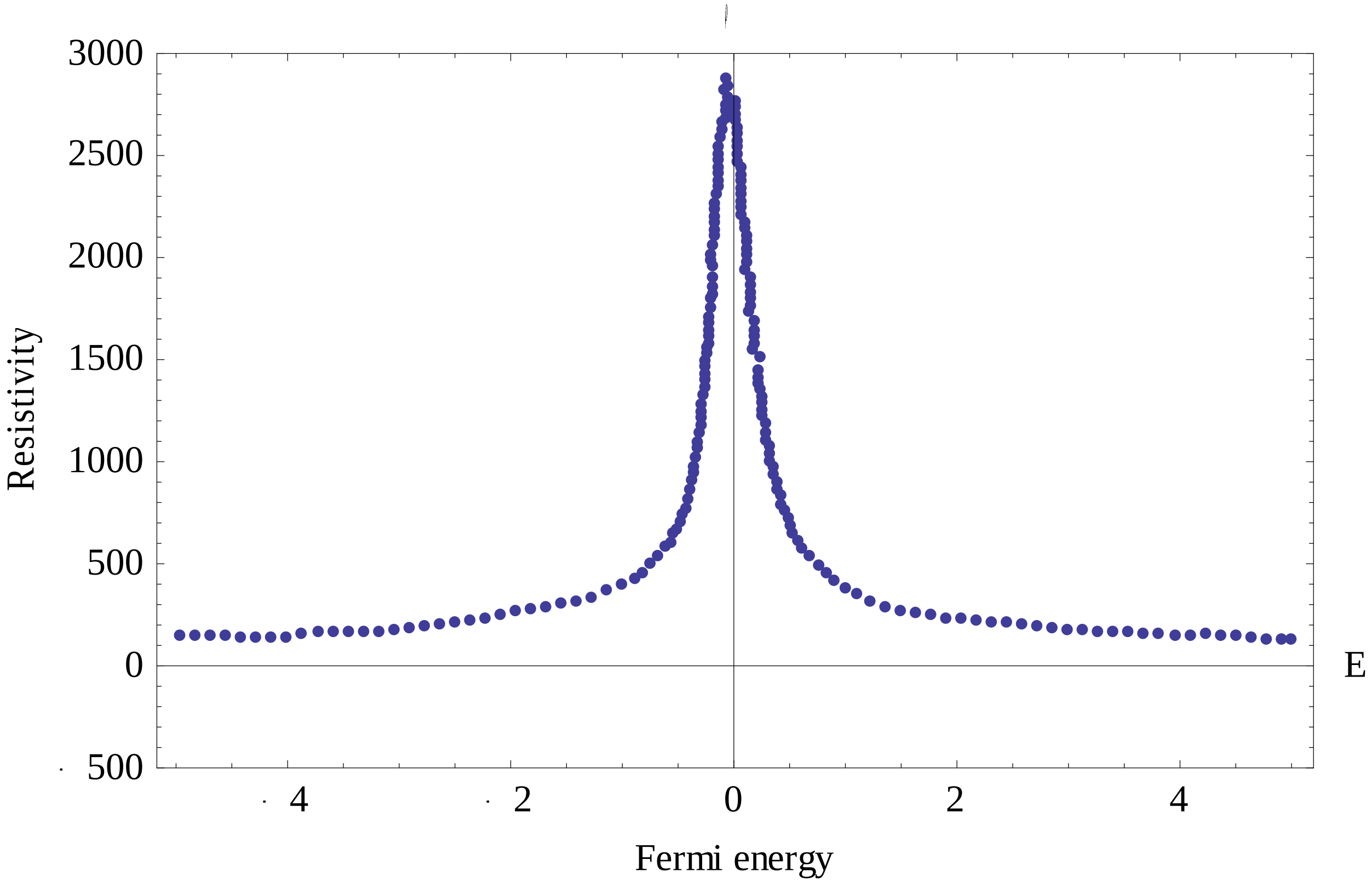

**Fig. 3**

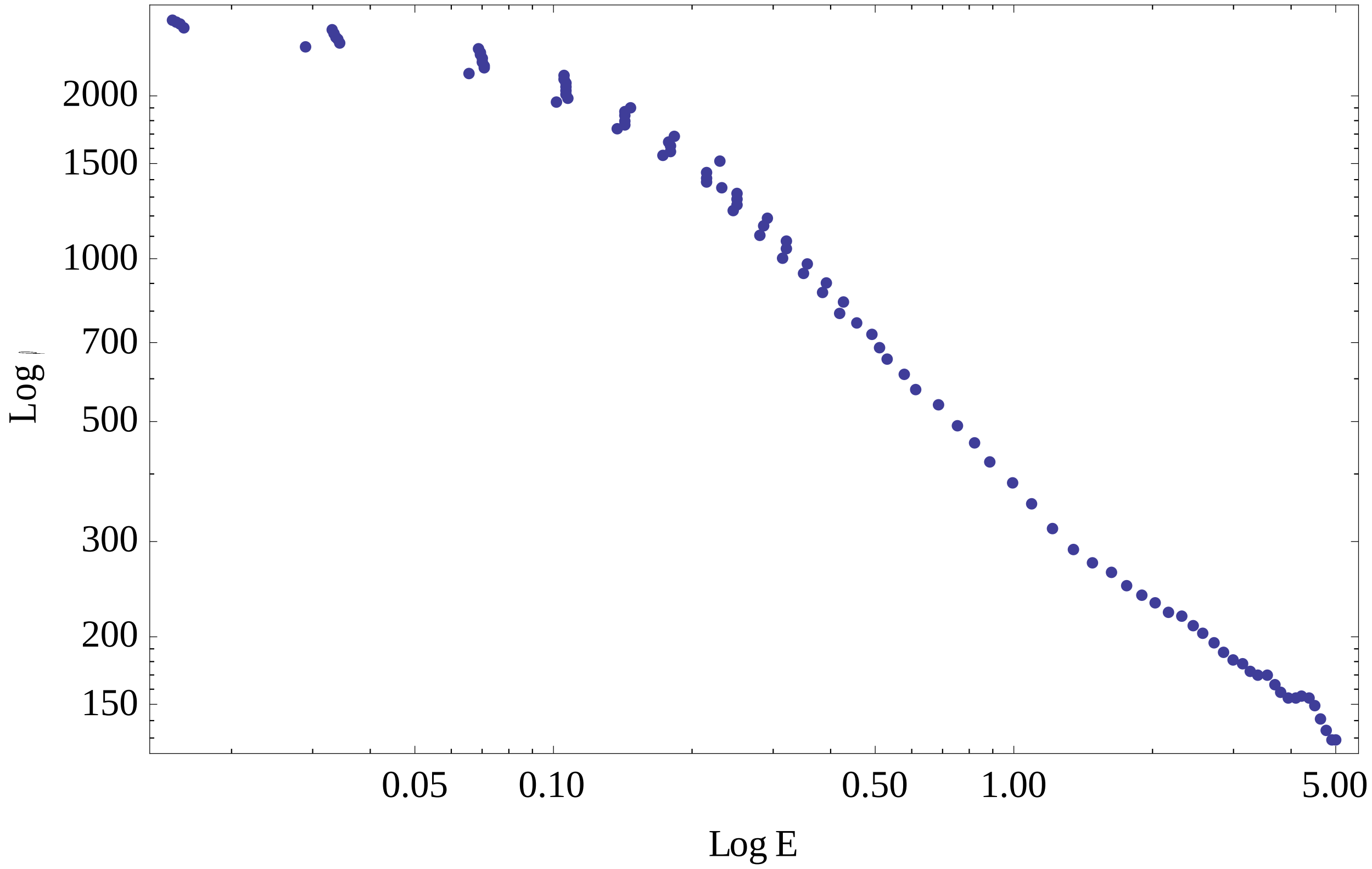

**Fig. 4**